# AN OVERVIEW OF TEXT-TO-SPEECH SYSTEMS AND MEDIA APPLICATIONS

**ABSTRACT**
Producing synthetic voice, similar to human-like sound, is an emerging novelty of modern interactive media systems. Text-To-Speech (TTS) systems try to generate synthetic and authentic voices via text input. Besides, well known and familiar dubbing, announcing and narrating voices, as valuable possessions of any media organization, can be kept forever by utilising TTS and Voice Conversion (VC) algorithms . The emergence of deep learning approaches has made such TTS systems more accurate and accessible. To understand TTS systems better, this paper investigates the key components of such systems including text analysis, acoustic modelling and vocoding. The paper then provides details of important state-of-the-art TTS systems based on deep learning. Finally, a comparison is made between recently released systems in term of backbone architecture, type of input and conversion, vocoder used and subjective assessment (MOS). Accordingly, Tacotron 2, Transformer TTS, WaveNet and FastSpeech 1 are among the most successful TTS systems ever released. In the discussion section, some suggestions are made to develop a TTS system with regard to the intended application.

## 1. INTRODUCTION

Text-to-speech (TTS) is an essential component of many speech-enabled applications such as navigation systems, and accessibility for the visually impaired [1]. Typical modern text-to-speech systems are complex to design [2]. These systems usually need a text frontend to extract linguistic features, a model to predict acoustic features and a signal-processing-based vocoder to reconstruct the final waveform [2] as shown on **Figure 1**. Each of these blocks require expert design and needs to train independently [2]. Neural network based TTS has achieved higher quality compared to conventional concatenative and statistical parametric approaches [3]. Concatenative synthesis mostly relies on concatenation of separated pieces of speech stored in the database [4]. Concatenative synthesis can generate high intelligibility speech close to original voice but it requires huge amount of data to match all possible combination of speech units, but on the other hand, such systems generate voices with lower naturalness quality [4] due to the less flexibility of concatenation of emotional and prosody features. To address such drawbacks of concatenative systems, statistical parametric synthesis speech (SPSS) is proposed. In fact, instead of directly producing an output waveform, acoustic features are firstly extracted, and then final waveform is reconstructed through such predicted parameters [4]. The advantages of SPSS are the output naturalness together with the flexibility to modify the parameters and therefore, creae audio with different prosody. However, the generated speech still suffers from low intelligibility and robotic-like artifacts [4]. Some early neural network modeling follow the same architecture as SPSS but replacing each block with its corresponding neural network like Deep Voice 1/2 [1, 5]. Later, one type of end-to-end (E2E) models like Tacotron 1/2 [2, 6], Deep Voice 3 [7], FastSpeech 1/2 [3, 8], which take character/phoneme sequences directly as input and produce Mel-spectrograms as acoustic features and finally utilise well-known vocoders to generate final waveform. Other types are fully E2E models developed to directly generate waveform from input character/phoneme sequences like FastSpeech 2s [8], ClariNet [9] and EATS [10].

## 2. TTS AND MEDIA

TTS can be also used in different media applications and can help develop media products cheaper, faster and more easily. Automatic news generation by the voice of a special announcer at regular times obviates the need for human voice costs. Moreover, a high-quality TTS can play a narrator's role in filmmaking and eases the reproduction of documentaries and middle programme clips. China Central Television [11], Japan Broadcasting Corporation (NHK) [12] and British Broadcasting Corporation (BBC) R&D [13] have recently investigated the TTS following an AI anchor (announcer).

## 3. KEY COMPONENTS OF A TEXT-TO-SPEECH SYSTEM

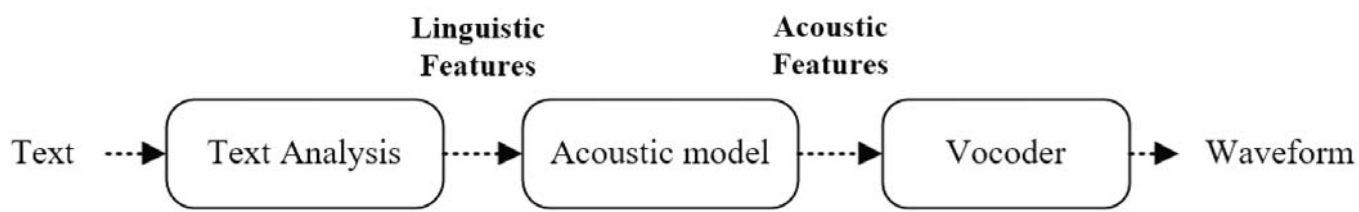

**Figure 1**: General Structure of TTS systems [4]

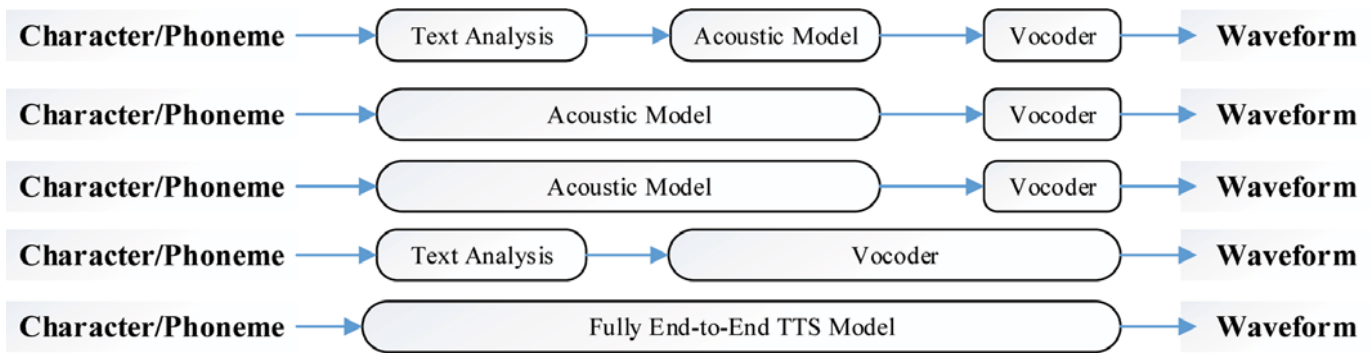
*Figure 2: Different types of TTS model [4]*

### TEXT ANALYSIS
Text analysis, also called frontend in TTS systems, transforms input text into linguistic features containing rich information about pronunciation and prosody [4]. Char2Wav and DeepVoice1/2 incorporate the character-to-linguistic feature conversion purely based on neural networks [4]. The prosody information, such as rhythm, stress, and intonation of speech corresponds to the variations in syllable duration, loudness and pitch [4]. A manually collected grapheme-to-phoneme lexicon is usually leveraged for G2P conversion.

### ACOUSTIC MODELS
Acoustic features are generated from linguistic features or directly from phonemes and characters. Some important acoustic features are Mel-Cepstral Coefficients (MCC), Mel-generalised coefficients (MGC) [355], band aperiodicity (BAP), fundamental frequency (F0), Voiced/Unvoiced (V/UV), Bark-Frequency Cepstral Coefficients (BFCC), and the most widely used Mel-spectrograms. In contrast to traditional acoustic models, neural based TTS systems mostly take character or phoneme sequence as input and generate high-level Mel-spectrograms as output [4]. An important advantage of neural end-to-end acoustic models over conventional ones is how alignment is done. In fact, conventional acoustic models require alignment between linguistic and acoustic features, while sequence-to-sequence based neural models implicitly learn the alignment through attention, or predict the duration jointly [4].

### VOCODER
Generally speaking, vocoders are classified into traditional SPSS and neural-based ones. STRAIGHT [14] and WORLD [15] are popular SPSS. Traditional vocoders generally take acoustic features as input while neural-based vocoders can take both acoustic and linguistic features. For example, WaveNet [16], WaveRNN [17] and char2wave take linguistic features and LPCNet [18], WaveGlow [19]and MelGAN [20] take acoustic features.

### 4. WHY END-TO-END TTS SYSTEMS?
An end-to-end TTS means to train the model merely by <text, audio> pairs. Such systems make the entire structure easier by removing the need to design each block independently. Another advantage of such systems is the flexibility to condition different types of input. For example, you can replace <text, audio> with <specific feature, audio> such as <MFCC, audio>, <Mel-spectrogram, audio> etc. Moreover, such systems are more robust compared with multi-stage ones. Therefore, these advantages imply that an end-to-end system could allow us to utilise more of huge real-world data [2].

It is worth noting the slight difference between an integrated end-to-end and vanilla end-to-end; the vanilla end-to-end structure is a compound of separated blocks trained independently and connected together to form the end-to-end system, like Deep Voice [1], while in integrated synthesisers, there is just one structure trained from scratch with random initialisation like Tacotron [2].

### 5. CHALLENGES TO END-TO-END TTS SYSTEM
In TTS systems, a given text input can lead to different pronunciation or speaking style output. This makes end-to-end TTS more difficult than other simple end-to-end tasks. Moreover, unlike automatic speech recognition (ASR) and Neural Machine Translation (NMT), TTS outputs are continuous and usually have much longer sequences. Each of these properties can cause errors which accumulate quickly [2].

*Figure 2* shows different types of TTS system. Type 1 is based on traditional structure including text analysis, acoustic model and vocoder (SPSS). Type 2 and type 4 combines text analysis and acoustic model and directly predicts general acoustic features or Mel-Spectrogram respectively using input characters. Tacotron 2, DeepVoice 3 and FastSpeech 1/2 are such examples. Type 3 TTS combines acoustic models and final vocoder, which directly transforms linguistic features into waveform (WaveNet). Finally, fully E2E models combine all three blocks into one and directly converts characters into waveform (Wave-Tacotron, Char2Wav, ClariNet).

### 6. DETAIL OF SOME IMPORTANT STATE-OF-THE-ART TTS
#### 3.1 WAVENET
Inspired by generative autoregressive models in image [21, WaveNet {Oord, 2016 #78}] addresses the same issue for generating wideband raw audio waveforms with at least 16,000 samples per second. WaveNet adopts dilated causal convolutions to deal with long-range temporal dependencies [16]. This architecture, when conditioned on a specific speaker, could apply to applications such as voice conversion, text-to-speech etc. The main ingredient of WaveNet is causal convolution [16]. In dilated convolutions, a fixed-sized filter covers larger area by skipping input values by certain steps. Dilated convolution allows the network to sweep more time-steps and therefore, modelling more temporal resolution. A stack of dilated casual convolutional layers is illustrated in the *Figure 3*.

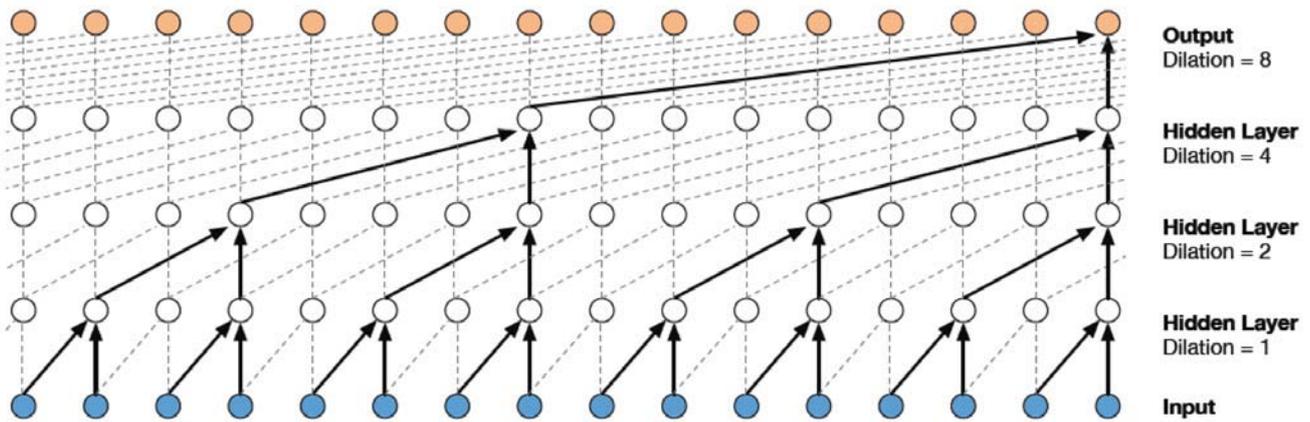

*Figure 3: structure of dilated convolution layer [16]*

**Table 1:** subjective results of WaveNet [16]

| Speech samples | Subjective 5-scale MOS in naturalness | |
|---|---|---|
| | North American English | Mandarin Chinese |
| LSTM-RNN parametric | 3.67 ± 0.098 | 3.79 ± 0.084 |
| HMM-driven concatenative | 3.86 ± 0.137 | 3.47 ± 0.108 |
| **WaveNet** (L+F) | **4.21 ± 0.081** | **4.08 ± 0.085** |
| Natural (8-bit μ-law) | 4.46 ± 0.067 | 4.25 ± 0.082 |
| Natural (16-bit linear PCM) | 4.55 ± 0.075 | 4.21 ± 0.071 |

One can condition the WaveNet on global and local parameters. The global conditioning occurs as a latent representation embedded. Local conditioning happens when the input time series in typically lower than output, thus, the first step is to up-sample the input (using transposed convolution network) to map it to the same resolution as the output and feed it fed into the process. However, the inputs to WaveNet such as linguistic features, predicted log fundamental frequency (F0), and phoneme durations require elaborate text-analysis [6] and exact prior processing. Due to its flexibility to generate audio, it can tackle a wide range of speech applications such as music, speech enhancement, voice conversion, source separation etc. [16].

### 3.2 DEEP VOICE 1

Deep voice is based on traditional text-to-speech pipelines and adopts the same structure but replaces each component with its corresponding neural network [1, 2]. According to Deep Voice, TTS systems consist of five major building blocks [1] :

- *The grapheme-to-phoneme model*
  Grapheme-to-phoneme (G2P) conversion is the process of generating pronunciation for words based on their written form. It has a highly essential role for natural language processing, text-to-speech synthesis and automatic speech recognition systems [22]. Deep Voice G2P model is based on an encoder-decoder architecture [1]. **Figure 4** illustrates the bi-directional LSTM[23].

- *The segmentation model*
  Given an audio file and phoneme-by-phoneme transcription of the audio, the segmentation module finds the start and time point of each phoneme. Adopting a state-of-the art speech recognition system could help to segment and label the given phoneme sequence and corresponding audio utterance [1]. This block adopts a CTC-based RNN.

- *The phoneme duration model*
  Predicting the temporal duration of each phoneme in an utterance. The model architecture used here is shared with fundamental frequency model and explained in the following part.

- *The fundamental frequency model*
  Having predicted voiced phonemes, this component predicts the fundamental frequency (f0) throughout its duration identified in the past section. Adopting a joint structure to predict phoneme duration and fundamental frequency, Deep Voice suggests an architecture comprising two fully connected layers with 256 units each followed by two unidirectional recurrent layers with 128 GRU cells each and finally a fully connected output layer [1]. The final layer outputs three estimations for every input phoneme: the phoneme duration, the probability that the phoneme is voiced, and F0 sampled uniformly over the predicted duration [1].

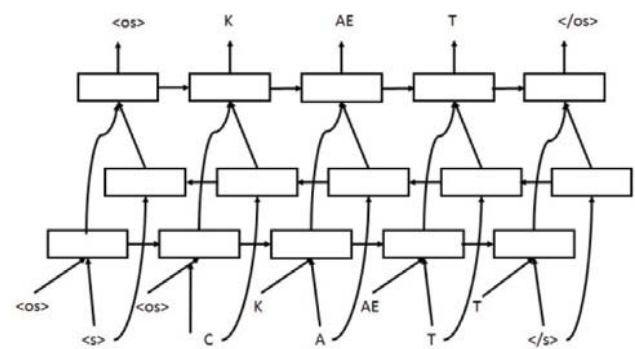

*Figure 4: The bi-directional LSTM. "CAT" for forward amd "TAC" for backward [1]*

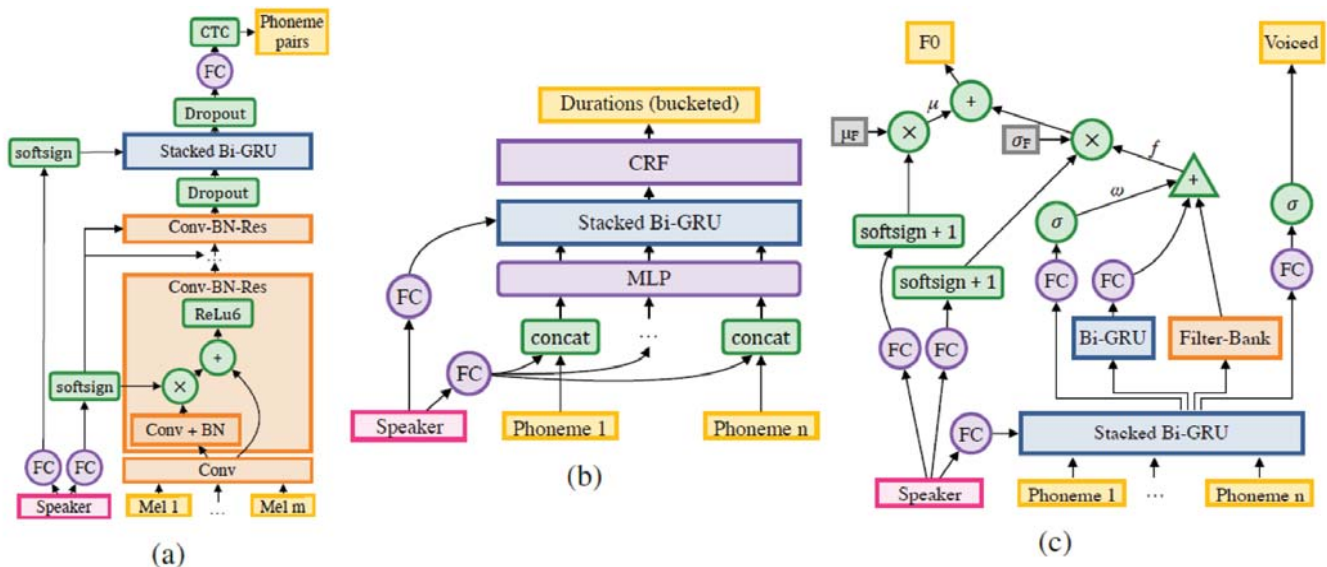

*Figure 5: Speaker embedding at each block of Deep Voice 2: a) segmentation b) duration c) frequency model [5]*

- *The audio synthesis model*
Combining the outputs of grapheme-to-phoneme, duration, and fundamental frequency prediction models, this module synthesises the output raw waveform. Deep Voice utilises WaveNet [16] as its backbone of audio synthesis.

In fact, Deep Voice has upgraded each block with the corresponding neural network based models. This model has not meaningfully progressed past the state-of-the-art systems in this regard. It concludes that the main barrier to progress towards natural TTS lies with duration and fundamental frequency prediction [1].

### 3.3 DEEP VOICE 2
This model keeps the general structure of the Deep Voice 1. Major differences with Deep Voice 1 are the separation of the phoneme duration and frequency models. The phoneme durations are predicted first and are then used as inputs to the frequency model. All models are trained separately. Each block is also augmented with low-dimensional speaker embedding. Using speaker-embedding technique in each block turns Deep Voice 2 into a multi-speaker TTS. Deep Voice 2 is a high quality text-to-speech systems and conclusively shows that neural speech synthesis models can learn effectively from small amounts of data spread among hundreds of different speakers [5].

### 3.4 DEEP VOICE 3
Deep voice 3 is fully convolutional character-to-spectrogram architecture TTS. It generates monotonic attention behaviour, avoiding error modes commonly affecting sequence-to-sequence models [7].

### 3.5 CHAR2WAV
Char2wave divides modelling text-to-speech models into two parts: a reader (frontend) and a neural vocoder (backend). The reader is an encoder-decoder architecture with attention that takes as input text or phonemes and transforms it into linguistic features. The vocoder accepts linguistic (acoustic) features and produces the corresponding audio [24]. In this definition, WaveNet [16] is considered as a neural backend [16]. Char2wav adopts SampleRNN as hierarchical vocoder. Char2Wav integrates the frontend and backend and learn the entire model end-to-end and simultaneously [24]. The general structure of Char2Wav is illustrated in the **Figure 6**. However, Char2Wav need to predict vocoder features firstly and, SampleRNN vocoder used here needs to be separately pre-trained whereas Tacotron directly predicts output raw waveform[2].

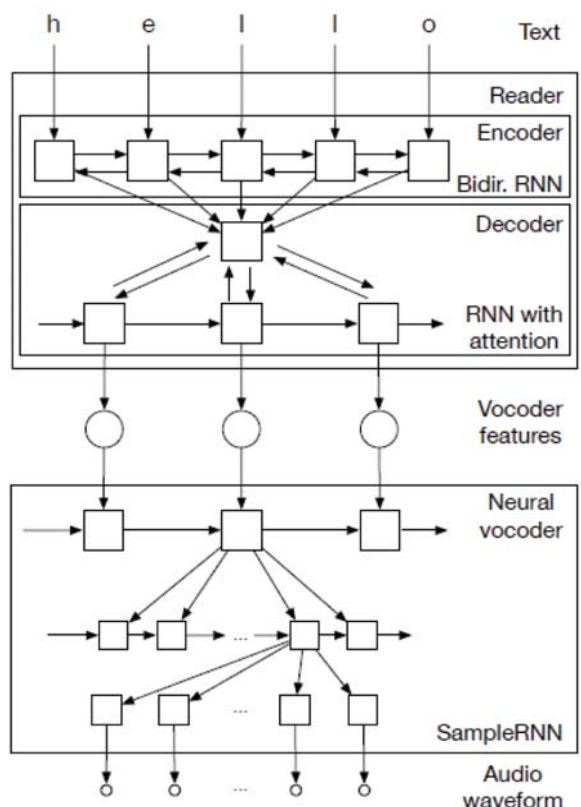

*Figure 6: Char2Wav: An end-to-end speech synthesis model [24]*

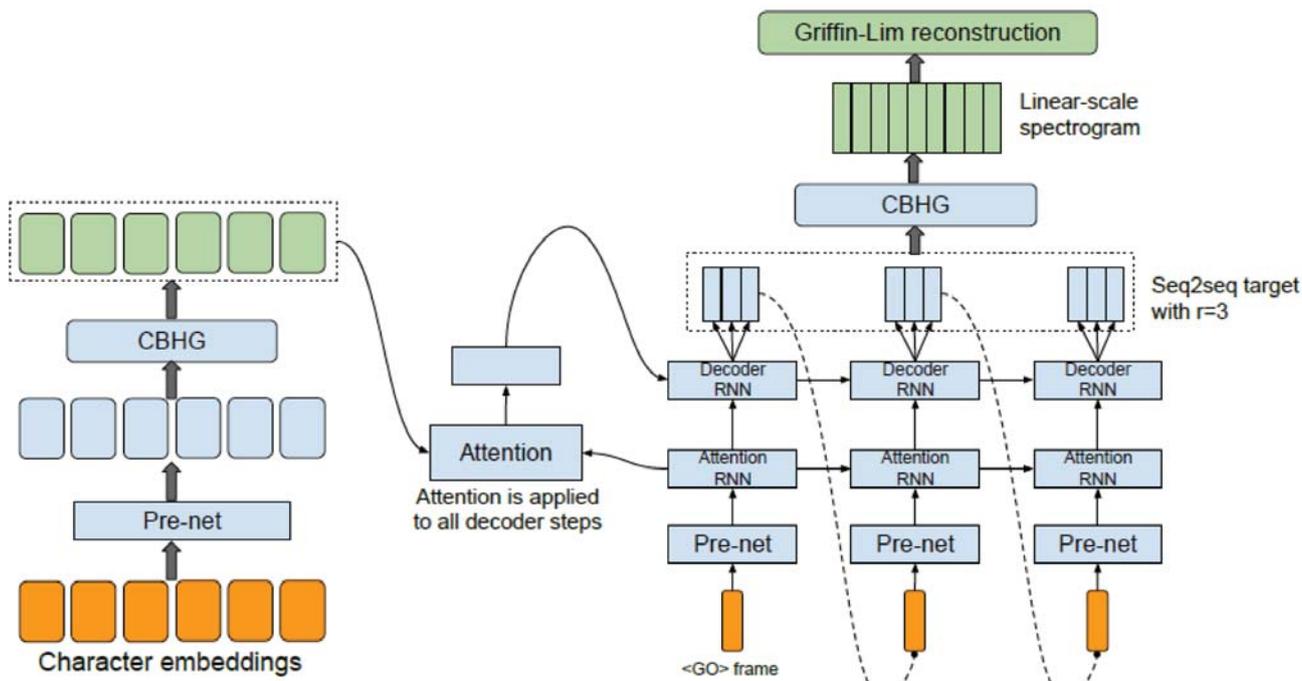

*Figure 7: Tacotron architecture [2]*

### 3.6 TACOTRON

Tacotron is an end-to-end generative text-to-speech model that directly synthesises speech from characters [2]. Tacotron is a sequence-to-sequence architecture producing a magnitude spectrogram from a sequence of characters which replaces traditional linguistic and acoustic features pipeline with a single neural network [6]. Tacotron consists of three blocks: an encoder, an attention-based decoder and a post-processing net. Tacotron takes characters as input and produces spectrogram frames, which are then converted to waveforms using Griffin-Lim. CBHG (1-D Convlution bank + Highway network + Bidirectional GRU) module of encoder reduces overfitting [2]. **Figure 7** illustrates the structure. **Table 2** shows the Mean Score Opinion (MOS) of Tacotron.

**Table 2:** MOS evaluation of Tacotron

| **Tacotron** | 3.82±0.085 |
|---|---|
| **Parametric** | 3.69±0.109 |
| **Concatenative** | 4.09±0.119 |

### 3.7 TACOTRON 2

Unlike Tacotron, which uses Griffin-Lim algorithm to produce final audio waveform, Tacotron 2 utilises WaveNet [16] as vocoder to compensate for the typical artifacts made by Griffin-Lim. As shown in figure 5, Tacotron 2 consists of two general blocks: (1) a recurrent sequence-to-sequence feature prediction network with attention, which predicts frames of mel-spectrograms and (2) a WaveNet vocoder to produce final audio waveform [6]. Tacotron 2 significantly outperforms all other TTS systems, and results in an MOS comparable to that of the ground truth audio [6].

### 3.8 TRANSFORMER TTS NETWORK

Transformer TTS network [25] is based on Tacotron 2 [6] and Transformer network [26]. Inspired by the success of Transformer networks in Neural Machine Translation (NMT), this approach suggests the use of Transformer combined with Tacotron 2 to predict mel-spectrograms and leveraging WaveNet to reconstruct final waveform. Since WaveNet is used to reconstruct the final raw waveform, this model is autoregressive and still suffers from slow inference [25]. According to the results reported in the paper, this approach outperforms Tacotron 2 with a gap of 0.048 and is very close to human quality (4.39 vs 4.44 MOS) [25]. The structure of Transformer TTS is illustrated in **Figure 8**.

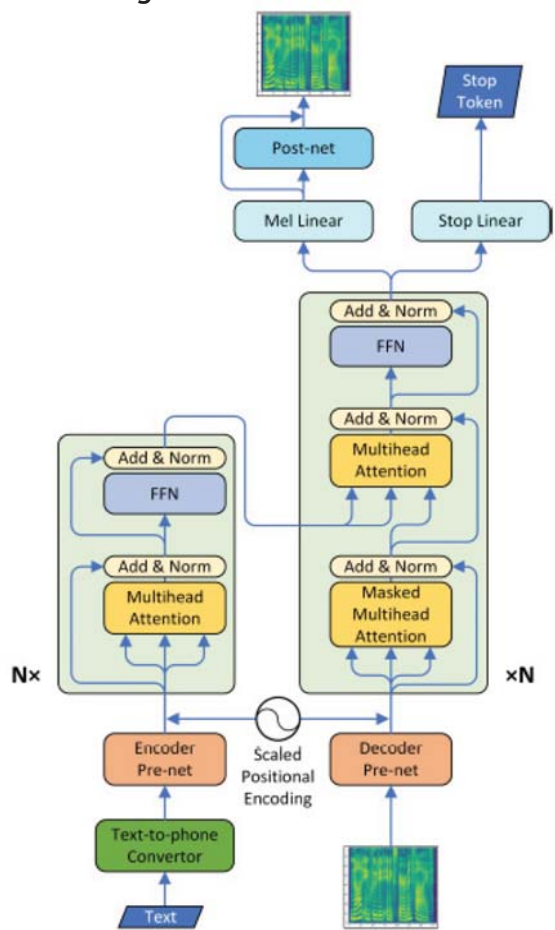

*Figure 8: Architecture of Transformer-TTS [25]*

## 7. DISCUSSION & CONCLUSION

**Table 3** collates the state-of-the-art TTS systems. Accordingly, Tacotron 2, Transformer TTS, WaveNet, FastSpeech 1 and ClariNet are among the most successful TTS systems ever released. In the discussion section, some suggestions are made to develop a TTS system with regard to the intended application. Some TTS models are end-to-end and convert the input characters to high-level acoustic features such Mel-spectrogram and adopts a neural vocoder such as WaveNet to generate output waveform. Others are fully end-to-end; converting input character/phoneme sequence directly to output waveform. Char2Wav, ClariNet and FastSpeech 2s are such examples. Therefore, enough notice is to be taken to differentiate between E2E and fully E2E. It is worth noting that, typically, the first versions of TTS systems are trained on single-speaker datasets such as LJSpeech, Internal US English and later versions are designed and trained on multi-speaker datasets such as VCTK. Embedding speaker vector to different blocks of a TTS can customizse it to the specific speaker.

**Table 3**: Comparison of State-Of-The-Art TTS models

| NAME | TYPE | BACKBONE ARCHITECTURE | INPUT | OUTPUT | VOCODER | DATASET (HOUR) (S-M*) | REPORTED MOS |
|---|---|---|---|---|---|---|---|
| WaveNet [16] | Autoregressive | PixelCNN | Linguistic Features | Wav | - | North American English (24.6) (S) | 4.21±0.081 |
| | | | | | | Mandarin Chinese (34.8) (S) | 4.08±0.085 |
| Deep Voice 1 [1] | Autoregressive | CNN-based | Character / phoneme | Linguistic Features | WaveNet | Internal English Speech (20) (S) (Synthesized Duration and F0) | 2.00±0.23 |
| | | | | | | Subset of the Blizzard 2013 (20.5) | 2.67±0.37 |
| Deep Voice 2[5] | Autoregressive | CNN-based | Character / phoneme | Linguistic Features | WaveNet | VCTK (44) (M) | 3.53±0.12 |
| | | | | | | Audiobooks (238) (M) | 2.97±0.17 |
| Deep Voice 3 [7] | autoregressive | Fully CNN-based + attention + Seq2Seq | Character / phoneme | Acoustic Features | Griffin-Lim | VCTK (44) (M) | 3.01 ±0.29 |
| | | | | | WORLD | | 3.44±0.32 |
| | | | | | WaveNet | | - |
| CHAR2WAV [24] | Autoregressive | Seq2Seq RNN | Character / Phoneme | Wav | SampleRNN | - | - |
| Tacotron [2] | Autoregressive | RNN + Encoder-Decoder + Attention | Character / phoneme | Acoustic Features | Griffin-Lim | Internal North American English (24.6) (S) | 3.82±0.085 |
| Tacotron 2 [6] | Autoregressive | RNN + Encoder-Decoder + Attention | Character / phoneme | Acoustic Features | WaveNet | Internal North American English (24.6) (S) | 4.5±0.06 |
| Transformer TTS [25] | Autoregressive | Transformer-based | Character / phoneme | Acoustic Features | WaveNet | Internal US English (25) (S) | 4.39 |
| ClariNet [9] | Autoregressive | CNN-based | Character / phoneme | Wav | WaveNet | Internal English speech dataset (20) | 4:15 ± 0:25 |
| Fastspeech [3] | Non-autoregressive | Transformer-based | Character / phoneme | Acoustic Features | WaveGlow | LJSpeech (24) (S) | 3.84±0.08 |
| Fastspeech 2 [8] | Non-autoregressive | Transformer-based | Character / phoneme | Acoustic Features | Parallel WaveGAN | LJSpeech (24) (S) | 3.83 ± 0.08 |
| Fastspeech 2s [8] | Non-autoregressive | Transformer-based | Character / phoneme | Wav | Parallel WaveGAN | LJSpeech (24) (S) | 3.71 ± 0.09 |

- *S stands for single speaker and M stand for multi-speaker.
- *Single architecture to jointly predict phoneme duration and time-dependent fundamental frequency.*

### ABOUT AUTHOR

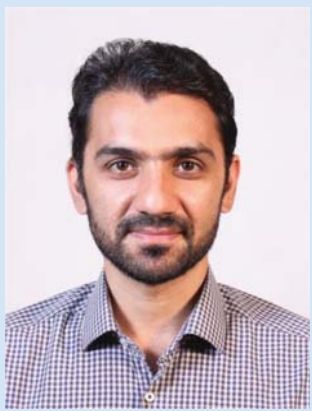

**Mohammad Reza Hasanabadi** received his B.S. in electrical engineering from Ferdowsi University of Mashhad in 2013, and his M.S. in Sound Engineering from IRIB University in 2017.

He is currently a Ph.D. candidate in telecommunication engineering at Shahid Beheshti University (SBU), Tehran, IRAN.

In 2019, he joined IRIB R&D. His research interests include deep learning and AI audio video fields, particularly focusing on speech processing, voice conversion, text-to-speech, audio coding, and their applications to broadcast systems. ■